\pdfoutput=1

\documentclass[11pt,twoside,a4paper,cmspaper,final,collab]{cms-tdr}
\def\svnVersion{394845}\def\cmsCernNoTag{CERN-EP/2016-286}\def\cmsCernDate{\today}\def\cmsMessage{Published in the Journal of High Energy Physics as \href{http://dx.doi.org/10.1007/JHEP03(2017)077}{\doi{10.1007/JHEP03(2017)077.}}}

\begin{document}\cmsNoteHeader{EXO-16-016}

\hyphenation{had-ron-i-za-tion}
\hyphenation{cal-or-i-me-ter}
\hyphenation{de-vices}
\RCS$Revision: 394750 $
\RCS$HeadURL: svn+ssh://svn.cern.ch/reps/tdr2/papers/EXO-16-016/trunk/EXO-16-016.tex $
\RCS$Id: EXO-16-016.tex 394750 2017-03-17 17:11:19Z florez $
\newlength\cmsFigWidth
\ifthenelse{\boolean{cms@external}}{\setlength\cmsFigWidth{0.85\columnwidth}}{\setlength\cmsFigWidth{0.4\textwidth}}
\ifthenelse{\boolean{cms@external}}{\providecommand{\cmsLeft}{top\xspace}}{\providecommand{\cmsLeft}{left\xspace}}
\ifthenelse{\boolean{cms@external}}{\providecommand{\cmsRight}{bottom\xspace}}{\providecommand{\cmsRight}{right\xspace}}
\providecommand{\cPj}{\ensuremath{\mathrm{j}}\xspace}
\providecommand{\cN}{\ensuremath{\mathrm{N}}\xspace}
\providecommand{\ST}{\ensuremath{S_\mathrm{T}}\xspace}
\renewcommand{\PW}{\ensuremath{\mathrm{W}}\xspace}
\cmsNoteHeader{EXO-16-016}
\title{\texorpdfstring{Search for heavy neutrinos or third-generation leptoquarks in final states with two hadronically decaying $\tau$ leptons and two jets in proton-proton collisions at $\sqrt{s} = 13\TeV$}{Search for heavy neutrinos or third-generation leptoquarks in final states with two hadronically decaying tau leptons and two jets in proton-proton collisions at sqrt(s) = 13 TeV}}

\date{\today}

\abstract{
A search for new particles has been conducted using events with two high transverse momentum (\pt) $\tau$
leptons that decay hadronically, at least two high-\pt jets, and missing transverse energy from the $\tau$ lepton decays.
The analysis is performed using data from proton-proton collisions, collected by the CMS experiment in 2015 at $\sqrt{s} = 13$\TeV, corresponding to
an integrated luminosity of 2.1\fbinv.
The results are interpreted in two physics models.
The first model involves heavy right-handed neutrinos, $\cN_{\ell}$ $(\ell = \Pe,\mu,\tau)$, and right-handed charged bosons, $\PW_\mathrm{R}$, arising in a left-right
symmetric extension of the standard model. Masses of the $\PW_\mathrm{R}$ boson below
2.35 (1.63)\TeV are excluded at 95\% confidence level, assuming the $\cN_{\tau}$ mass is 0.8 (0.2) times the mass of the $\PW_\mathrm{R}$ boson and that
only the $\cN_{\tau}$ flavor contributes
to the $\PW_\mathrm{R}$ decay width. In the second model, pair production of third-generation scalar leptoquarks that decay into $\tau\tau\PQb\PQb$ is considered. Third-generation scalar leptoquarks with masses below 740\GeV are excluded, assuming a 100\% branching fraction for the leptoquark decay to a $\tau$ lepton and a bottom quark.
This is the first search at hadron colliders for the third-generation Majorana neutrino, as well as the first search for third-generation leptoquarks in the final state with a pair of hadronically decaying $\tau$ leptons and jets.
}

\hypersetup{
pdfauthor={CMS Collaboration},%
pdftitle={Search for heavy neutrinos or third-generation leptoquarks in final states with two hadronically decaying tau leptons and two jets in proton-proton collisions at sqrt(s) = 13 TeV},%
pdfsubject={CMS},%
pdfkeywords={CMS, physics, heavy neutrinos, leptoquarks}}

\maketitle
\section{Introduction}\label{sec:intro}
In this paper a search for new phenomena at the CERN LHC beyond the standard model (SM) of particle physics is presented, using events containing
two energetic $\tau$ leptons and at least two energetic jets. This final state is expected, for example, in  the decay of a
right-handed \PW boson ($\PW_\mathrm{R}$) into a $\tau$ lepton and a heavy neutrino that decays into another
$\tau$ lepton and two jets ($\tau\tau \cPj\cPj$). The same final state is also expected from the decay of leptoquark (LQ) pairs.
A brief description of the two models predicting these different decay paths to the same final state is given below.

In the SM, the neutrinos of the three generations are considered to be massless,
while the observation of neutrino oscillations implies otherwise.
One way to generate neutrino masses is the seesaw mechanism,
which can be accommodated in a left-right symmetric extension of the SM (LRSM)~\cite{Lindner:2001hr, Minkowski:1977sc, Mohapatra:1979ia}.
This model explains the observed parity violation in the SM as the consequence of spontaneous symmetry breaking at a multi-\TeV
mass scale and introduces a right-handed counterpart to the SM group SU(2)$_{\mathrm{L}}$.
The new SU(2)$_\mathrm{R}$ gauge group is associated with three new
gauge bosons, $\PW_\mathrm{R}^{\pm}$ and Z', and three heavy right-handed neutrino states $\cN_{\ell}$ ($ \ell = \Pe,\mu,\tau$),
partners of the light neutrino states $\nu_{\ell}$. A reference process allowed by this model is
the production of a $\PW_\mathrm{R}$ that decays into a heavy neutrino $\cN_{\ell}$ and a charged lepton of the same generation.
The heavy neutrino subsequently decays into a lepton and two jets.

In this context, the light neutrino mass is given by $m_{\nu} \sim y_{\nu}^{2}v^{2} / m_{\cN}$, where $y_{\nu}$ is a Yukawa coupling to the SM Higgs field,
$v$ the
Higgs field vacuum expectation value in the SM, and $m_{\cN}$ the mass of the heavy neutrino state.
In type I seesaw models, the light and heavy neutrinos must be Majorana particles in order to explain the known neutrino
masses. As a consequence, processes that violate lepton number conservation by two units would be possible. Therefore, searches for heavy Majorana
neutrinos can provide important tests of the nature of neutrinos and the origin of neutrino masses.

A similar dilepton plus dijet final state can be realized in other extensions of the SM that predict scalar or vector LQs.
The motivation for postulating such particles is to achieve a unified description of quarks and leptons \cite{GlashowGUT}.
Leptoquarks are SU(3) color-triplet bosons that carry both lepton and baryon numbers \cite{SalamLQ,PhysRevD.11.703.2,Gripaios:2009dq}, and are foreseen in
grand unified theories, composite models, extended technicolor models, and superstring-inspired models.
The exact properties (spin, weak isospin, electric charge, chirality of the fermion couplings, and fermion number)
depend on the structure of each specific model. For this reason, direct searches for LQs at collider experiments
are typically performed in the context of the Buchm{\"u}ller--R{\"u}ckl--Wyler model \cite{BRW}.
This model includes a general effective lagrangian describing interactions of LQs with SM fermions
and naturally provides symmetry between leptons and quarks of the SM.
Since they carry both baryon number and lepton number, it is expected that LQs would be produced in pairs, and that LQs
of the n${th}$ generation would decay into leptons and quarks of the same generation.

This analysis is a search for the third-generation particles of the LRSM and LQ models,
considering final states that contain a pair of $\tau$ leptons and jets.
In the heavy neutrino search, these final states arise from the decays
$\PW_\mathrm{R} \to \tau + \cN_{\tau} \to \tau + \tau \qqbar$.
This search is the first for the third-generation Majorana neutrino at hadron colliders.
Previous searches for heavy neutrinos have been performed at LEP~\cite{Abreu:1996pa,Adriani:1992pq}, excluding heavy neutrinos of this model for masses below approximately 100\GeV, and in the dimuon 
plus dijet ($\mu\mu \cPj\cPj$) and dielectron plus dijet ($\Pe\Pe \cPj\cPj$) channels at 7\TeV by ATLAS~\cite{ATLAS:2012ak}
and at 8\TeV by CMS \cite{Khachatryan:2014dka}. The ATLAS and CMS searches assumed that $\cN_{\tau}$ is too heavy to play a role in the
decay of $\PW_\mathrm{R}$.
In those searches, the $\PW_\mathrm{R}$ mass ($m(\PW_\mathrm{R})$) is excluded up to approximately 3.0\TeV.
In the search for LQs, requiring the presence of $\tau$ leptons selects third-generation LQs, leading to the final state $\tau\tau\PQb\PQb$.
Searches for LQs in this channel from ATLAS at 7\TeV \cite{LQ3Atlas} and CMS at 13\TeV \cite{LQ3CMS} excluded third generation
leptoquarks for masses less than 534\GeV and 740\GeV, respectively.

The $\tau$ lepton is the heaviest known lepton and decays about one third of the time into purely leptonic final states ($\tau_{l}$),
and the remainder of the time into hadrons plus one neutrino.
In this analysis pairs of $\tau$ leptons are selected in which both decay hadronically ($\tauh$) into one, three, or (rarely)
five charged mesons often accompanied by one or more neutral pions.
Because the hadronic decay of the $\tau\tau$ system has two associated neutrinos, the events have missing transverse momentum
($\ptvecmiss$), where $\ptvecmiss$ is defined as the negative vector sum of the transverse momenta of all reconstructed particles in an event.
The magnitude of $\ptvecmiss$ is referred to as \MET.
In contrast to heavy neutrino searches in the $\Pe\Pe\cPj\cPj$ or $\mu\mu \cPj\cPj$ final states, this analysis uses events that
contain neutrinos from the tau lepton decays,
and thus the $\PW_\mathrm{R}$ resonance cannot be fully reconstructed in the $\tauh\tauh$ channel.
To distinguish between signal and SM processes that give rise to a similar final state topology (backgrounds),
the visible $\tau$ lepton decay products, two jets, and the $\MET$ are used to reconstruct the partial mass:
\begin{linenomath}
\begin{equation}
\small{
   m(\tau_{ \mathrm{h},1},\tau_{ \mathrm{h},2},\cPj_{1},\cPj_{2},\MET) = \sqrt{(E^{\tau_{ \mathrm{h},1}}+E^{\tau_{ \mathrm{h},2}}+E^{\cPj_{1}}+E^{\cPj_{2}}+\MET)^{2}-
   (\overrightarrow{p}^{\tau_{ \mathrm{h},1}}+\overrightarrow{p}^{\tau_{ \mathrm{h},2}}+\overrightarrow{p}^{\cPj_{1}}+\overrightarrow{p}^{\cPj_{2}}+\ptvecmiss)^{2}},
}
\label{eq:ditaudijetmass}
\end{equation}
\end{linenomath}
where $E$ and $p$ represent the energies and momenta of selected $\tau$ and jet candidates.

The partial mass is expected to be large in the heavy neutrino case and close to the $\PW_\mathrm{R}$ mass.
The heavy neutrino search strategy is to look for a broad enhancement in the partial mass distribution inconsistent with known SM backgrounds.
For pair production of leptoquarks, the scalar sum of the transverse momenta (\pt) of the decay products,
$\ST=\pt^{\tau_{ \mathrm{h},1}}+\pt^{\tau_{ \mathrm{h},2}}+\pt^{\cPj_{1}}+\pt^{\cPj_{2}}$, is expected to be large
and comparable with the total leptoquark mass.
In this case the strategy is similar to other leptoquark analyses and involves searching for a broad enhancement in the high-\ST part of the spectrum.

It is worth noting that the partial mass and \ST are typically
higher than in channels containing $\tau_{l}$, because of the different number of neutrinos from $\tau$ lepton decays.
At the same time, because a $\tauh$ resembles a jet, the typical probability of misidentifying a jet as a $\tauh$ is at least an order of magnitude higher than that for a jet to be misidentified as an electron or muon. As a result, the multijet background from quantum chromodynamics (QCD) processes in the $\tauh\tauh$ channel is larger than in the $\tau\tau \to \tau_{l}\tauh$ and
$\tau\tau \to \tau_{l}\tau_{l}$ channels.
However, the QCD multijet contribution at high values of partial mass and \ST is
strongly reduced owing to its rapidly decreasing production cross section.
These considerations, combined with the fact that the considered final state has the highest branching fraction to $\tauh$ pairs, makes it a promising channel in searches for new physics.

The analysis is performed using proton-proton collision data collected by the CMS experiment in 2015 at $\sqrt{s} = 13$\TeV. The overall strategy is similar to the previously cited heavy neutrino and leptoquark searches. Upon selecting two high-quality $\tauh$ candidates and two additional jet candidates, the distribution of $m(\tauh,\tauh,\cPj,\cPj,\MET)$ or
\ST is used to look for a potential signal that would appear as an excess of events over the SM expectation at large
values of the mass or \ST.
The object reconstruction is described in Section \ref{sec:leptonRecoId}, followed by the description of the signal and background simulation samples in Section \ref{sec:backgrounds}.
The selections defining the signal region (SR), described in Section~\ref{sec:eventSelections}, achieve a reduction of the background
to a yield of about 1 event in the region where signal dominates. A major challenge of this analysis is to ensure the signal and trigger efficiencies are not only
high, but well understood. This is accomplished through studies of SM processes involving genuine $\tauh$ candidates.
The analysis strategy is described in Section~\ref{sec:estimation} and relies on the selection of $\Z(\to \mu\mu$)+jets and
$\Z\to \tauh \tauh$ events.
A number of background enriched control regions are defined in Section~\ref{sec:estimation}.
The purpose of the control samples is to ensure a good understanding of the background contributions as well as to cross-check the accuracy of the efficiency
measurements and assign appropriate systematic uncertainties (Section~\ref{sec:systematics}). Estimates of the background contributions
in the SR are derived from data wherever possible, using samples enriched with background events.
These control regions are used to measure the partial mass shapes, \ST shapes, and selection efficiencies in order to extrapolate to the region where the signal is expected.
In cases where the background contributions are small ($<$10\%) or the above approach is not feasible, data-to-simulation scale factors,
defined as a ratio between the numbers of observed data events and expected simulated yields in background-enhanced regions, are used to validate or correct the expected contributions obtained from the simulation samples. Finally, the results are discussed in Section \ref{sec:results}.
\section{The CMS detector}
\label{sec:cmsdet}
A detailed description of the CMS detector, together with a definition of the coordinate system
used and the relevant kinematic variables, can be found in \cite{CMS}.

The central feature of the CMS apparatus is a
superconducting solenoid of 6\unit{m} inner diameter, providing a field
of 3.8\unit{T}. Within the field volume are the silicon pixel and strip
tracker, the lead tungstate crystal electromagnetic calorimeter (ECAL),
which includes a silicon sensor preshower detector in front of the ECAL endcaps,
and the brass and scintillator hadron calorimeter.
In addition to the barrel and endcap detectors, CMS has extensive forward
calorimetry. Muons are measured in gas-ionization detectors embedded in the steel flux-return yoke
of the solenoid.

The inner tracker measures charged particles within the region of pseudorapidity $\abs{\eta} < 2.5$
and provides an impact parameter resolution of $\sim$15mum\ and a
\pt resolution of about 1.5\% for particles with $\pt = 100\GeV.$ Collision events are selected by a first-level trigger composed of custom hardware processors
and a high-level trigger that consists
of a farm of commercial CPUs running a version of the offline
reconstruction optimized for fast processing.
\section{Object reconstruction and identification}\label{sec:leptonRecoId}

Jets are reconstructed
using the particle-flow (PF) algorithm~\cite{CMS-PAS-PFT-10-001}.
In the PF approach, information from all subdetectors is combined to reconstruct and identify final-state particles
(muons, electrons, photons, and charged and neutral hadrons) produced in the beam collisions.
The anti-\kt clustering algorithm~\cite{antikt} with a distance parameter of $R  =  0.4$ is used for jet
clustering.
Jets are required to pass identification criteria designed to reject particles from additional beam collisions within the same or a nearby
bunch crossing (pileup) and from anomalous behavior of the calorimeters, and to ensure separation from any identified
leptons by $\Delta R = \sqrt{\smash[b]{(\Delta\eta)^{2} + (\Delta\phi)^{2}}} > 0.4$, where $\phi$ is the azimuthal angle.
For jets with $\pt > 30$\GeV and $\abs{\eta} <2.4$,
the identification efficiency is approx 99\%, with a rejection efficiency of 90--95\% for jets originating from pileup interactions~\cite{CMS-PAS-JME-13-005}.
The jet energy scale and resolution are calibrated using correction factors that depend on the \pt and $\eta$ of the
jet \cite{CMS:JetResol}. Jets originating from the hadronization of bottom quarks are identified using the loose working point of the combined
secondary vertex algorithm \cite{Chatrchyan:2012jua}, which exploits observables related to the long lifetime of b hadrons.
For \PQb quark jets with $\pt > 30$\GeV and $\abs{\eta} <2.4$, the identification efficiency is
approximately 85\%, with a mistag rate of about 10\% for light-quark and gluon jets~\cite{CMS-PAS-BTV-15-001}.
The b quark jets are used to obtain $\ttbar$-enriched
control samples in order to estimate the background rate in the SR.

Although muons are not used to define the SR, they are used to obtain control samples for the background estimates.
Muons are reconstructed using the tracker and muon detectors.
Quality requirements based on the minimum number of hits in the silicon
tracker, pixel detector, and muon chambers are applied to suppress
backgrounds from decays in flight and from hadron shower remnants that reach the muon system \cite{Chatrchyan:2012xi}.
The muon identification efficiency for the quality requirements and kinematic range used in this analysis is approximately 98\%.
Muon candidates are additionally required to pass isolation criteria.
Isolation is defined as the \pt sum of the reconstructed PF charged and neutral particles, within an isolation cone of radius
$\Delta R =0.4$ centered around the muon track~\cite{muon13TeV}. The contribution from the muon candidate
is removed from the sum and corrections are applied to remove the contribution from particles produced in pileup interactions.

Hadronic decays of the $\tau$ lepton are reconstructed and identified using the ``hadrons-plus-strips" algorithm \cite{tes}
designed to optimize the performance of $\tauh$ reconstruction by including specific $\tauh$
decay modes. To suppress backgrounds from light-quark or gluon jets, a $\tauh$ candidate is
required to be isolated from other energy deposit in the event.
The isolation criterion is defined as the scalar \pt sum $S_{\tau}$ of charged and neutral
PF candidates within a cone of radius $\Delta R = 0.5$ around the $\tauh$ direction, excluding the
$\tauh$ candidate. The isolation criterion is $S_{\tau} < 0.8$\GeV.

Additionally, $\tauh$ candidates are separated from electrons and muons by using dedicated discriminators in the event.
The algorithm to discriminate a $\tauh$ from an electron uses observables that quantify the compactness and shape of energy deposits in the ECAL, in combination with observables that are sensitive to the amount of bremsstrahlung emitted along the leading
track and observables that are sensitive to the overall particle multiplicity. The discriminator against muons is based on the presence of hits in the muon system associated with the track of the $\tauh$ candidate.
The resulting combined efficiency for the isolation and selection requirements used to define the SR is 55\% averaged over the
kinematic range used in this analysis.

The presence of neutrinos in the $\tau\tau$ decays must be inferred from the imbalance of transverse momentum measured in the detector.
Information from the forward calorimeter is included in the calculation of $\MET$, and the
jet energy corrections described above are propagated as corrections to $\MET$.
Missing transverse energy is one of the most important observables for discriminating the signal events from background events that do not
contain neutrinos, such as QCD multijet events with light-quark and gluon jets.
\section{Signal and background simulation}\label{sec:backgrounds}

The QCD multijet processes are the dominant background in the SR.
Multijet events are characterized by jets with high multiplicity, which can be misidentified as a $\tauh$.
Apart from QCD multijets, the other much smaller backgrounds are the top pair production ($\ttbar$) and the Drell--Yan (DY) process giving rise to $\tau$ leptons plus jets.
The DY+jets events are characterized by two isolated $\tau$ leptons and additional jets from initial-state radiation.
Backgrounds from $\ttbar$ events
contain two b quark jets and either a genuine isolated $\tauh$ lepton or, with similar probability,
a misidentified $\tauh$ candidate.

Collision data are compared to samples of Monte Carlo (MC) simulated events and techniques based on control samples in data are employed
when possible. The \MADGRAPH (v5.1.5)\cite{Alwall:2011uj} program is used for simulation of DY+jets, \PW{}+jets,
and $\ttbar$+jets production at leading order.
The \MADGRAPH generator is interfaced with \PYTHIA 8 \cite{PYTHIA8}, for parton showering
and fragmentation simulation. The \PYTHIA generator is used to model the signal and QCD multijet processes.
The heavy-neutrino signal event samples are generated
with $\PW_\mathrm{R}$ masses ranging from 1 to 3\TeV. The $\cN_{\tau}$ mass varies between 0.05 and 0.95 multiplied by the $\PW_\mathrm{R}$ mass.
It is assumed that the gauge couplings associated with the left and right SU(2)
groups are equal and that the $\cN_{\tau}$ decays are prompt. 
It is also assumed the $\textrm{N}_{\textrm{e}}$ and $\textrm{N}_{\mu}$ masses are too heavy to play a role in the decay of $\textrm{W}_{\textrm{R}}$, and thus
$\textrm{W}_{\textrm{R}} \to \tau + \textrm{N}_{\tau}$ and $\textrm{W}_{\textrm{R}} \to q \bar{q}'$ are the dominant decay modes. The branching fraction for the
$\PW_{\textrm{R}} \to \tau + \mathrm{N}_{\tau}$ decay is approximately 10--15\%, depending on $m(\PW_{\mathrm{R}})$ and $m(\mathrm{N}_{\tau})$. 
The leptoquark signal event samples are generated with masses ranging from 200 to 1000\GeV.
The simulated events are processed with a detailed simulation of the CMS apparatus using
the \GEANTfour package \cite{Geant}.

In simulations, the DY and $\ttbar$ background yields, as well as the signal
yields, are normalized
to the integrated luminosity of the collected data using
next-to-leading order (NLO) or next-to-next-to-leading order (NNLO) cross sections \cite{Kramer:2004df, Hoeche:2014aia, PhysRevLett.115.062002, Czakon:2012zr}.
The mean number of interactions in a single bunch crossing in the analyzed data set is 21.
In simulated events, multiple interactions are superimposed on the primary collision,
with the distribution of the number of pileup interactions matching that observed in data.

\section{Event selection}\label{sec:eventSelections}

Candidate signal events were collected using a trigger requiring the presence of at least two $\tauh$ candidates
with $\pt > 35$\GeV and $\abs{\eta} < 2.1$ \cite{Zxsection}.
In addition to the requirements on $\tauh$ trigger objects, kinematic requirements on \pt and $\eta$
are imposed on the reconstructed $\tauh$ candidates used in the SR
to achieve a trigger efficiency greater than 90\% per $\tauh$ candidate. Events are required to have at least two $\tauh$ candidates with $\pt > 70$\GeV. The $\tauh\tauh$ pairs are required
to be separated by $\DR > 0.4$. Each $\tauh$ candidate is required to have $\abs{\eta} < 2.1$ in order to ensure that it is reconstructed fully within
the acceptance of the tracking system. Candidates are also required to satisfy the reconstruction and identification criteria described in
Section~\ref{sec:leptonRecoId}.
In contrast to other $\tau\tau$ analyses, an opposite-sign requirement cannot be used to discriminate against backgrounds from misidentified
$\tauh$ candidates, since the signal in the LRSM model can yield both oppositely-charged and same-sign
$\tauh\tauh$ pairs, because of the Majorana nature of the heavy neutrino.

In addition to the preselection described above, the final selection is defined by requiring at least two jets with $\pt > 50$\GeV
and $\abs{\eta} < 2.4$. Only jets separated from the $\tauh$ candidates by $\Delta R > 0.4$ are considered.
Because there are neutrinos in the $\tau\tau$ system decay, we are able to require $\MET > 50$\GeV to control the level of QCD multijet background.
Further, to reduce the contribution from \Z{}+jets events, the invariant mass of the $\tauh\tauh$ pair
is required to be $>$ 100\GeV.

The signal selection efficiency for $\PW_\mathrm{R} \to \tau + \cN_{\tau} \to \tau + \tau \qqbar$ events depends on the
$\PW_\mathrm{R}$ and $\cN_{\tau}$ masses. The total signal selection efficiency, assuming the $\cN_{\tau}$ mass is half the
$\PW_\mathrm{R}$ mass, is 1.65\% for $m(\PW_\mathrm{R})$ = 1.0\TeV and 5.15\% for
$m(\PW_\mathrm{R})$ = 2.7\TeV. The signal selection efficiency for LQ $\to\tau$b events is 4.14\% for $m(\mathrm{LQ})$ = 0.6\TeV and
6.68\% for $m(\mathrm{LQ})$ = 1.0\TeV.
These efficiencies include the branching fraction of approximately 42\% for $\tau\tau$ decaying to $\tauh\tauh$.
\section{Background estimation}\label{sec:estimation}

As discussed above, the $\MET$ and $\tauh$ isolation are the main variables discriminating against QCD multijet events.
The QCD multijet background estimation methodology utilizes control samples obtained by inverting the signal region
requirements on these two variables. In the remainder of this section, events obtained by inverting the isolation requirement on both
$\tauh$ candidates will be referred to as nonisolated $\tauh\tauh$
samples. The QCD multijet background is estimated using data and relying on the ``$ABCD$" method.
The regions $A$, $B$, $C$, and $D$ are defined as follows:

\begin{itemize}
  \item $A$: fail the $\MET > 50$\GeV requirement; nonisolated $\tauh\tauh$;
  \item $B$: fail the $\MET > 50$\GeV requirement; pass nominal isolation (as in SR);
  \item $C$: pass the $\MET > 50$\GeV requirement; nonisolated $\tauh\tauh$;
  \item $D$: pass the $\MET > 50$\GeV requirement; pass nominal isolation (as in SR).
\end{itemize}

Region $D$ is the nominal SR.
The QCD multijet components $N_{\mathrm{QCD}}^{i}$ in regions $i = A,B,$ and $C$ are predicted by subtracting simulated non-QCD backgrounds
from data ($N_{\mathrm{QCD}}^{i}=N_{\text{Data}}^{i}-N_{\text{non-QCD}}^{i}$). The signal contamination in the control regions is negligible according to simulation
($<$1\%). The contribution of QCD multijet events in the SR ($N_{\mathrm{QCD}}^{D}$) is estimated using the predicted rate of QCD multijet events
in region $C$ ($ N_{\mathrm{QCD}}^{C}$), weighted by a scale factor used to extrapolate from the nonisolated to the isolated $\tauh$ region.
The extrapolation factor is obtained by dividing the expected number of QCD multijet events in region $B$ ($N_{\mathrm{QCD}}^{B}$)
by the expected number of QCD multijet events in region $A$ ($N_{\mathrm{QCD}}^{A}$).
Therefore, the yield of QCD multijet events in the SR is given by $N_{\mathrm{QCD}}^{D} = N_{\mathrm{QCD}}^{C} \, (N_{\mathrm{QCD}}^{B} /
N_{\mathrm{QCD}}^{A})$.
The shapes for the variables of interest, $m(\tauh,\tauh,\cPj,\cPj,\MET)$ and \ST, are correlated with
$\MET$ and thus were obtained from region $C$.

\begin{figure}[tbh!]
  \centering
    \includegraphics[width=0.48\textwidth]{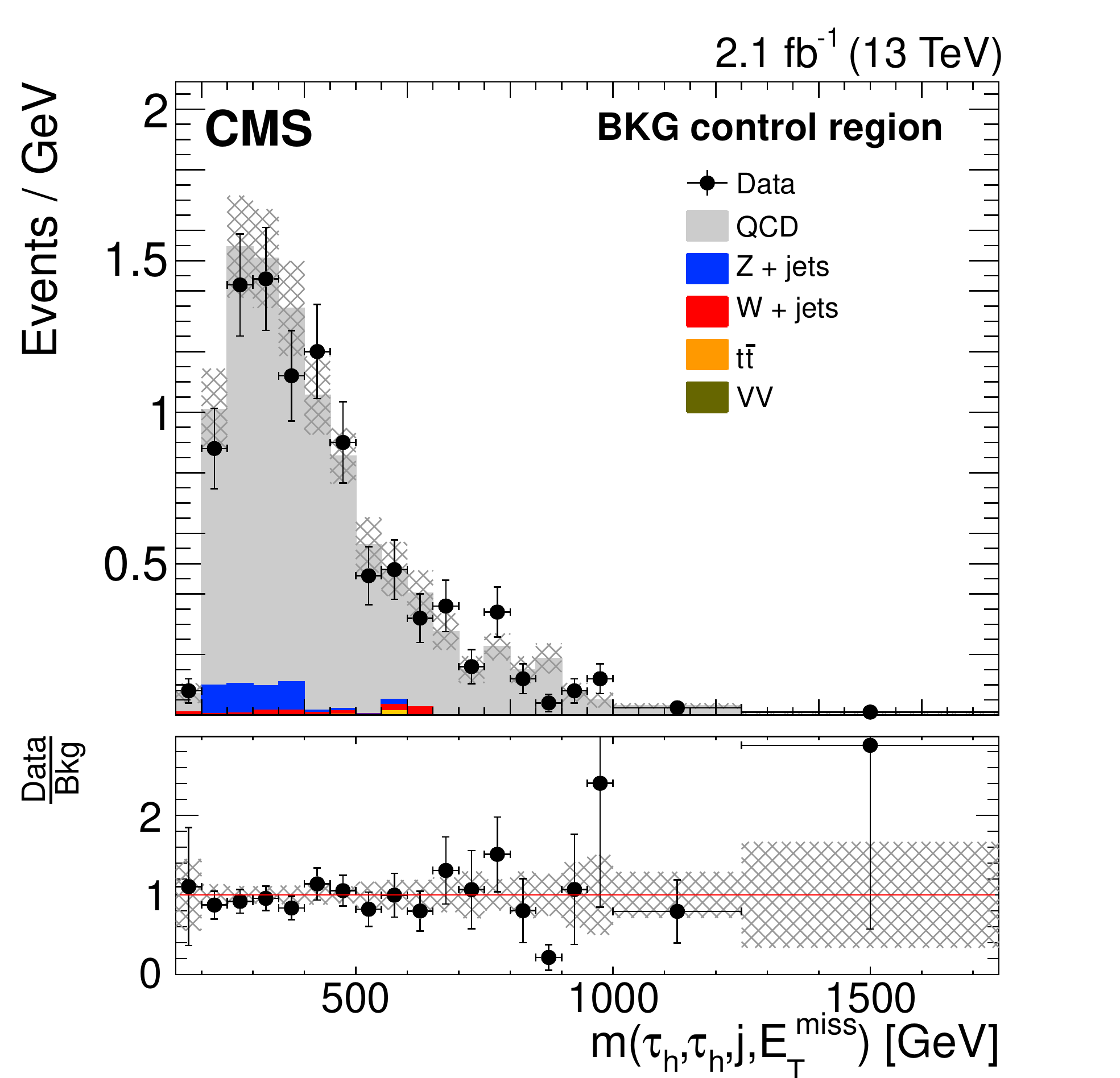}
    \includegraphics[width=0.48\textwidth]{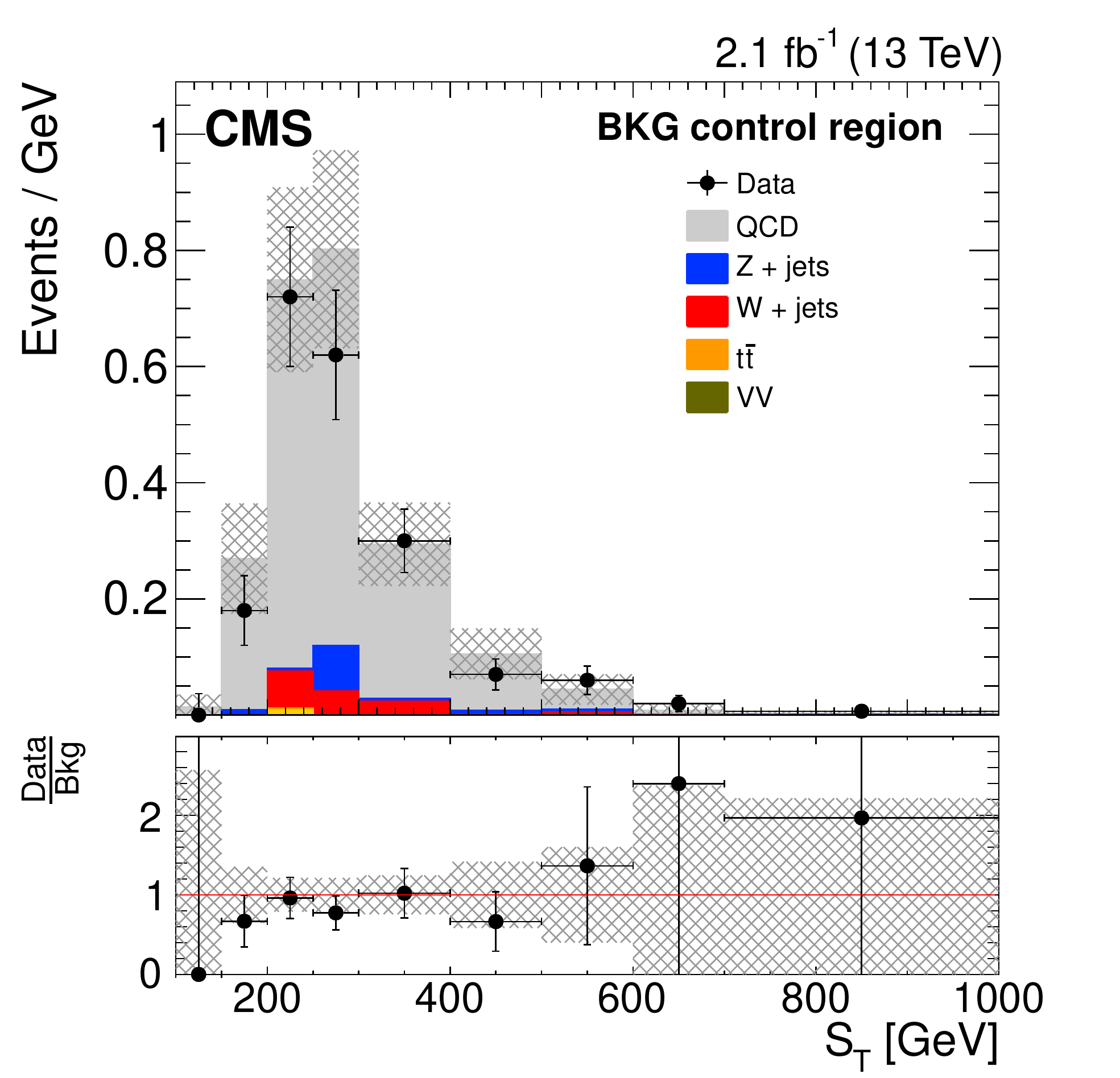} \\
    \includegraphics[width=0.48\textwidth]{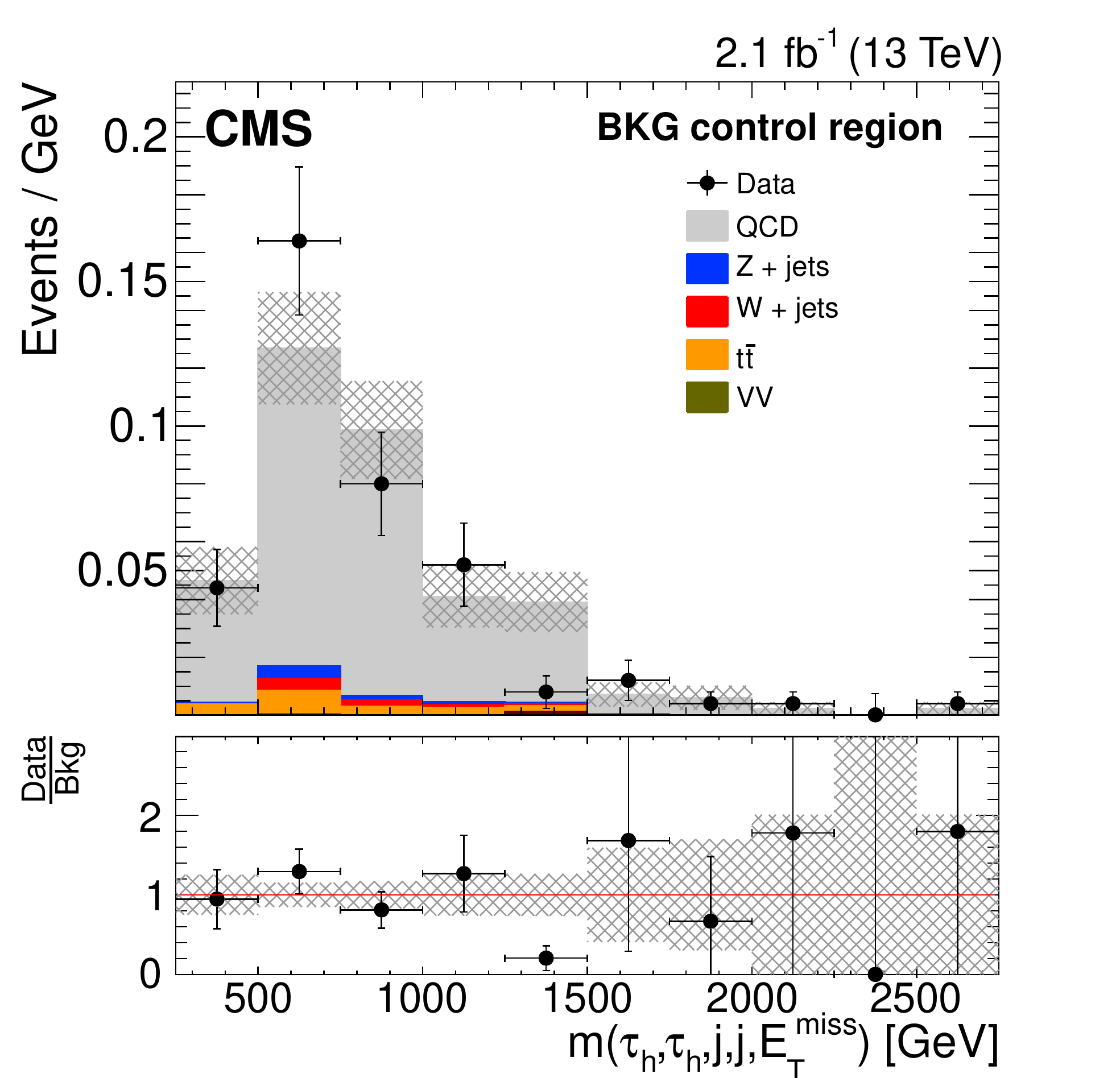}
    \includegraphics[width=0.48\textwidth]{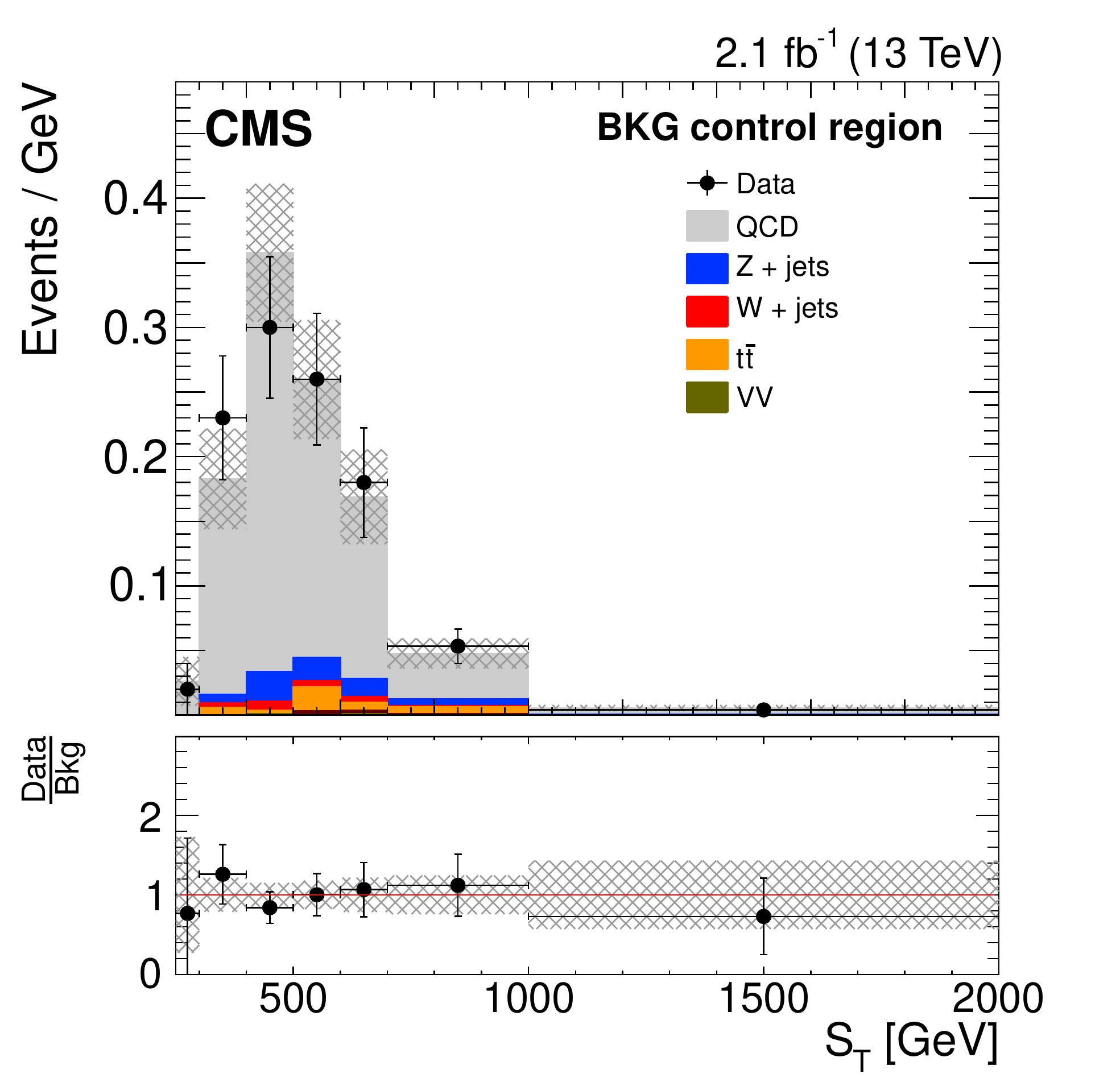}
\caption{Upper left: QCD multijet background shape validation test for $N_{\cPj}<2$, showing that the
$m(\tauh\tauh,\cPj,\ETmiss)$ distribution in the
nonisolated $\tauh\tauh$ control sample (``QCD" in the legend) correctly models the shape in the isolated region (``Data").
Upper right: QCD multijet background validation test of the $ABCD$ method applied to $\tauh\tauh$ data, showing that
there is a good agreement in the $S_\mathrm{T}$ distribution between the observed yield and shape and the predicted yield and shape.
Lower left: QCD multijet background validation test for $N_{\cPj} \geq 2$ data with $\MET < 50$\GeV, showing that the
$m(\tauh\tauh,\cPj,\cPj,\ETmiss)$ distribution
in the nonisolated $\tauh\tauh$ control sample correctly models the shape in the isolated region.
Lower right: QCD multijet background validation test for $N_{\cPj} \geq 2$ data, showing that the \ST
distribution in the nonisolated $\tauh\tauh$ control sample correctly models the shape in the isolated region.
The hatched band in the upper panel of each distribution represents the total statistical uncertainty in the background.
The lower panel shows the ratio between the observed data and the background estimation.
The shaded band in the lower panel represents the statistical uncertainty in the background prediction.
The diboson (``VV" in the legend) contributions are negligible.
}
  \label{fig:SignalSelectionPlots}
\end{figure}

Tests to validate both the normalization and shapes obtained for the background are performed with data.
The first set of validation tests in data is performed using the same method and event selection criteria described above for the different regions,
except with an inverted jet multiplicity requirement, $N_{\cPj} < 2$, in order to provide an exclusive set of regions, $A'$, $B'$, $C'$, and $D'$.
The purity of QCD multijet events in these control samples ranges approximately from 96 to 99\%.
There is agreement between the observation of $123$ QCD multijet events in region $D'$ and the prediction of $122.2 \pm 10.3$ events given
by the prescription $N_{\mathrm{QCD}}^{D'} = N_{\mathrm{QCD}}^{C'} \, (N_{\mathrm{QCD}}^{B'} / N_{\mathrm{QCD}}^{A'})$.
Figure~\ref{fig:SignalSelectionPlots} (upper) shows the $m(\tauh, \tauh, \cPj, \cPj, \MET)$ and \ST
distributions in region $D'$, where the shapes of QCD multijet events were obtained from region $C'$ and normalized to the
predicted yield of QCD multijet events in the region $D'$. There is agreement across the
$m(\tauh,\tauh,\cPj,\cPj,\MET)$ and \ST spectra, showing that $\tauh$ isolation does not bias
either distribution. An additional test on the extraction of the shape from the nonisolated $\tauh$ regions, with $N_{\cPj} \ge 2$,
is performed using the shape from QCD multijet events falling in region $A$, to estimate the shape of QCD multijet events in region $B$.
Figure~\ref{fig:SignalSelectionPlots} (lower) shows the $m(\tauh,\tauh,\cPj,\cPj,\MET)$ and \ST distributions in region $B$,
using the shape for QCD multijet events from region $A$, which provides further confidence in the method.
The procedure outlined in this section yields a QCD multijet estimate of $N_{\mathrm{QCD}}^{\mathrm{D}} = N_{\mathrm{QCD}}^{\mathrm{SR}} = 15.1 \pm 4.1$ events.
The overall uncertainty is dominated by the statistical uncertainty in the control samples.

The measurement of the $\Z(\to\tau\tau)$+jets contribution to the SR is
based on both simulation and data. The efficiency for the trigger and for the requirement of at least two high-quality $\tauh$ leptons is expected to be well
modeled by simulation. Mismodeling of the $\Z(\to\tau\tau)$+jets background rate and shapes in the SR can come from the requirement of two additional jets.
For these reasons we consider two control samples:
the first control sample is used to validate the correct modeling of the requirement of at least two high-quality $\tauh$ leptons;
the second control sample is used to measure a correction factor for the correct modeling of two additional jets.

The first $\Z(\to\tau\tau)$ control sample, used to validate the modeling of the trigger and the requirement of at least two high-quality
$\tauh$ leptons, is obtained by using the preselection requirements defined previously, and additionally requiring $\tauh\tauh$ pairs to have an invariant mass less than
100\GeV. This results in a sample composed of approx 90\% of $\Z(\to\tau\tau)$, according to simulation. The rates and shapes in data
and simulation are consistent, with a measured data-to-simulation scale factor of $0.97 \pm 0.19$.

The second control sample, used to measure a correction factor for the efficiency of the dijet selection, is obtained by applying criteria,
similar to those used in the search analysis, to select a sample of $\Z(\to \mu \mu)$+jets events having a dimuon invariant mass $m_{\mu\mu}$
compatible with that of the \Z boson ($60 < m_{\mu\mu} < 120$\GeV). Candidate events for this control sample were collected using a trigger
requiring the presence of at least one muon object with $\pt(\mu) > 18$\GeV. A study of this control sample allows a straightforward estimation,
using lepton universality in the \Z boson decay, of the extra hadronic activity expected in $\Z(\to\tau\tau)$+jets events.
Thus the rate for $\Z\to\tau\tau$ in simulation, after final selections, can be corrected using the measured dijet selection efficiency,
to determine the expected contribution of $\Z(\to\tau\tau)$+jets in the SR.
The measured correction factor is $1.20 \pm 0.01$, resulting in a DY+jets background estimate of $1.3 \pm 0.5$ events in the SR.
Systematic uncertainties in the estimated background yields are described in Section 7.

In a similar way, the estimation of the $\ttbar$ contribution to the SR is also obtained using information from both data and simulation.
A $\ttbar$-enriched control sample is obtained by applying all the signal selection criteria with at least one b-tagged jet and
two isolated muons, as opposed to $\tauh\tauh$, and additionally by requiring a \Z boson mass veto requirement
($m_{\mu\mu}$ outside the region between 80 and 110\GeV) to suppress $\Z(\to\mu\mu)$+jets background.
The $\ttbar$ prediction from simulation agrees with the observed yield and shape in the control sample.
The measured data-to-simulation scale factor in the control sample is $0.99 \pm 0.03$,
and thus the $\ttbar$ prediction in the SR is based on simulation, without any corrections.
The $\ttbar$ background yield in the SR is $2.5 \pm 0.9$ events.

\section{Systematic uncertainties}\label{sec:systematics}

Various imperfectly known or poorly simulated effects can alter the shape and normalization
of the $m(\tauh,\tauh,\cPj,\cPj,\MET)$ and \ST spectra.
Since the estimation of the background contributions in the SR is partly based on simulation,
the signal and certain backgrounds are affected by similar sources of systematic uncertainties. For example,
the uncertainty in the integrated luminosity measurement is 2.7\%~\cite{REFLUMI} and affects the signal and $\ttbar$ background.
The dominant sources of systematic uncertainties in the signal, DY+jets, and $\ttbar$ background predictions are the
uncertainties in the $\tauh$ identification and trigger efficiency.
The $\tauh$ trigger efficiency (the fraction of $\tauh$ candidates that additionally pass the $\tauh$ trigger
requirement) is estimated using a sample of $\Z\to\tau\tau\to \mu\tauh$ events, collected using a
single-muon trigger, that satisfy the same $\tauh$ identification criteria used to define the SR.
This estimation leads to a relative uncertainty of 5.0\% per $\tauh$ candidate.
Systematic effects associated with the $\tauh$ identification are extracted from a fit to the \Z$(\to\tau\tau)$
visible mass distribution, $m(\tau_1,\tau_2)$. In order to estimate the uncertainty in the $\tauh$ identification efficiency, the fit constrains the Z boson production cross section to
the measured cross section in the $\Z(\to \Pe\Pe/\mu\mu)$ decay channels,
leading to a relative uncertainty of 7\% per $\tauh$ candidate~\cite{Zxsection}.
An additional systematic uncertainty, which dominates for high-\pt $\tauh$ candidates,
is related to the confidence that the MC simulation correctly models the identification
efficiency. This additional uncertainty increases linearly with \pt and amounts to 20\%
per $\tauh$ candidate at $\pt = 1$\TeV.
The uncertainties related to the background estimation methods are negligible, as found from validation tests.
Additional contributions to the uncertainties in the signal, DY+jets, and $\ttbar$ background predictions
are due to the uncertainty in the $\tauh$/jet energy scale, ranging from 3--5\%.
The systematic uncertainty in the QCD multijet background prediction is dominated by the statistical uncertainty of the data
used in the control regions (about 27\%). The contamination from other backgrounds in the QCD multijet control regions has a negligible effect on the
systematic uncertainty.
The uncertainty in the signal acceptance due to the choice of parton distribution functions included in the simulated samples is
evaluated in accordance with the PDF4LHC recommendation and amounts to 5\%~\cite{Butterworth:2015oua}.
The systematic effect caused by imprecise
modeling of initial- and final-state radiation is determined  by reweighting events to account for effects such as missing $\alpha_{s}$
terms in the soft-collinear approach~\cite{softCollinear} and missing  NLO terms in the parton shower
approach~\cite{partonShower}.
The dominant effects that contribute to the $m(\tauh,\tauh,\cPj,\cPj,\MET)$ and \ST shape variations include
the $\tauh$ and jet energy scale uncertainties, resulting in systematic uncertainties of less than 10\% in all mass and \ST bins.
\section{Results}\label{sec:results}

\begin{figure}[tbh!]
  \centering
    \includegraphics[width=0.48\textwidth]{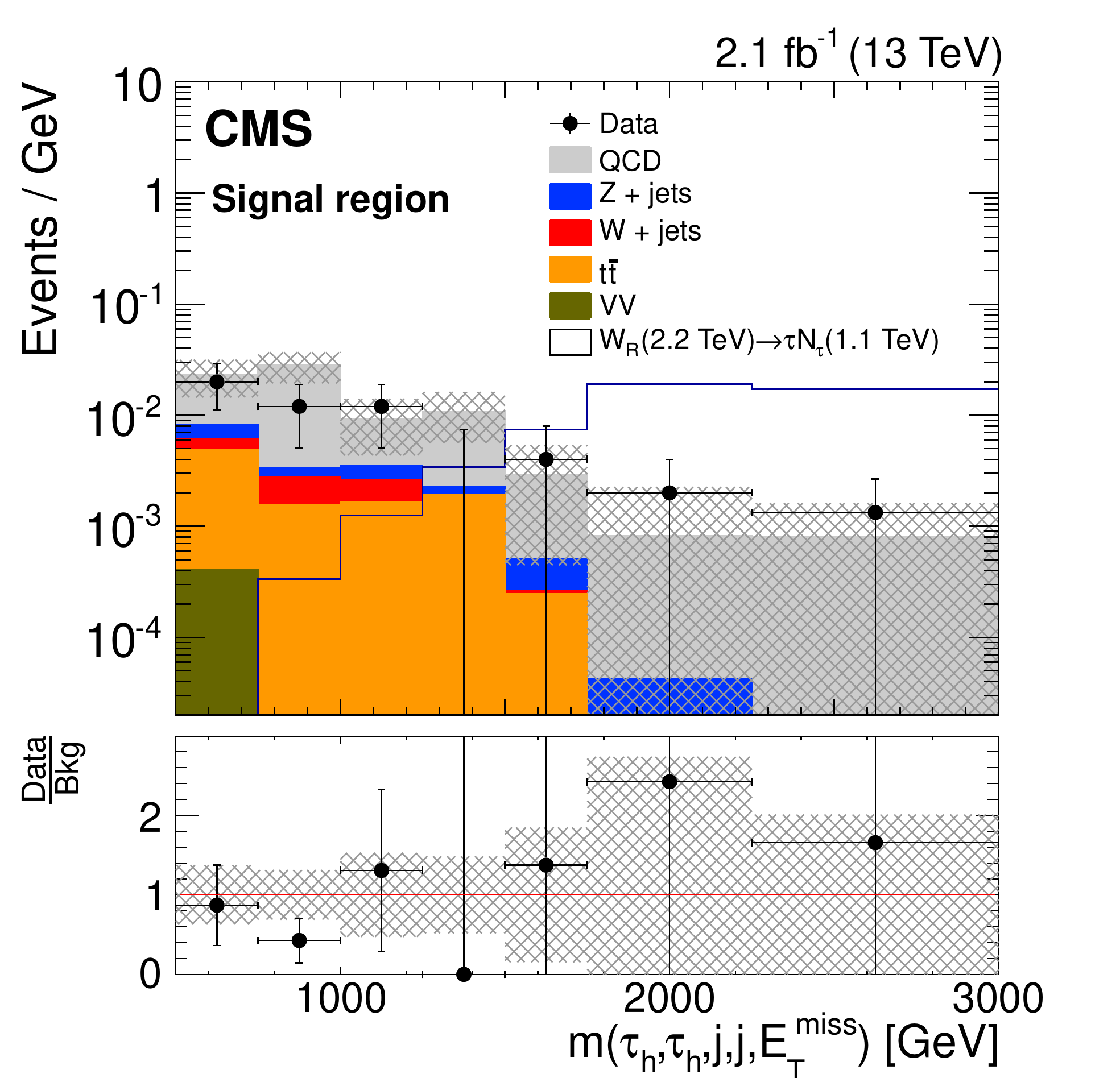}
    \includegraphics[width=0.48\textwidth]{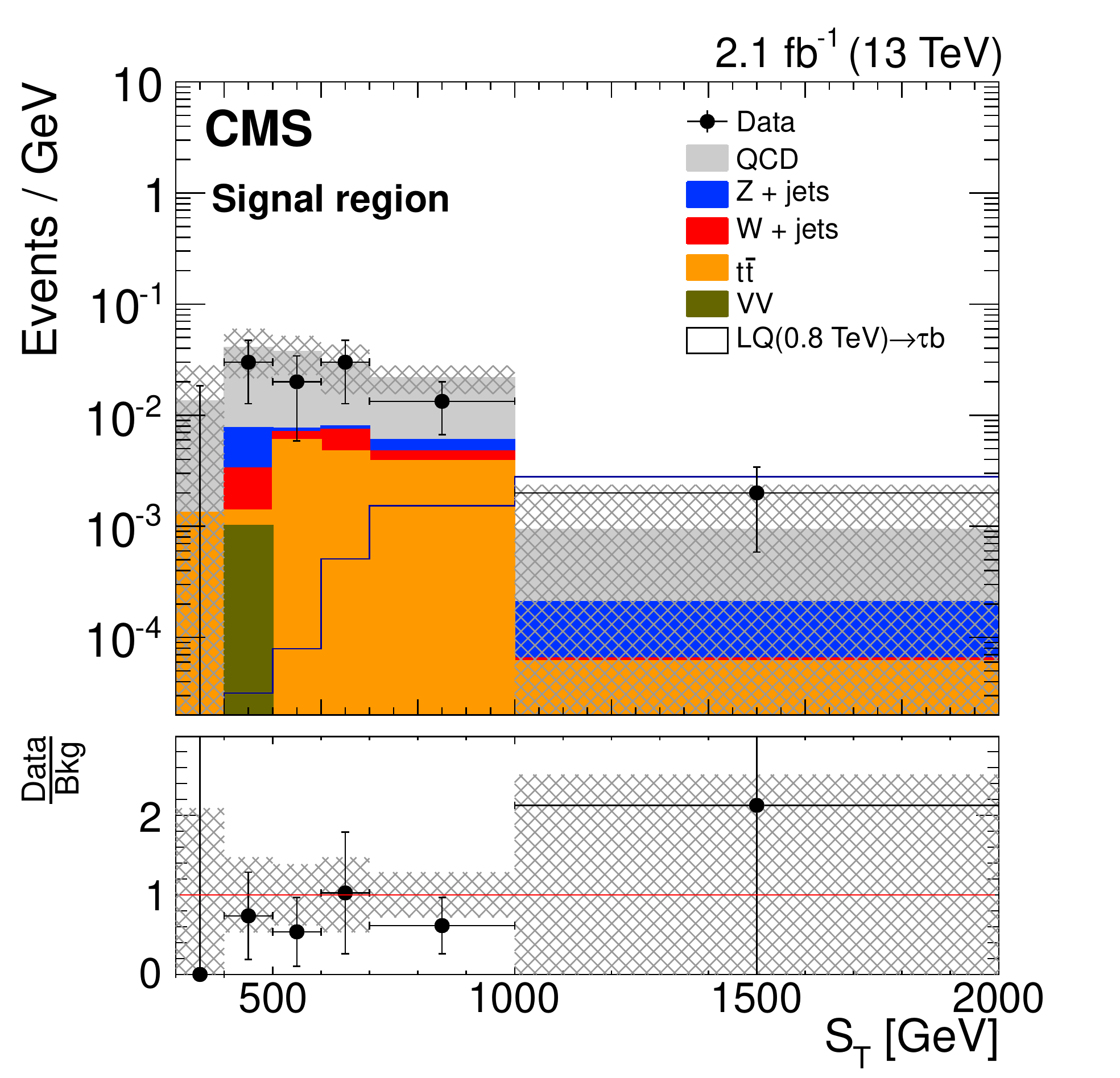}
  \caption{
Left:  $m(\tauh,\tauh,\cPj, \cPj, \MET)$ distribution in the SR.
Right: \ST distribution in the signal region. The estimated backgrounds are stacked while the data and simulated signal are overlaid.
The hatched band in the upper panel of each distribution represents the total statistical uncertainty in the background.
The lower panel shows the ratio between the observed data and the background estimation.
The shaded band across the lower panel, represents the total statistical and systematic uncertainty.
}
    \label{fig:SignalRegionPlot}
\end{figure}

Figure~\ref{fig:SignalRegionPlot} shows the background predictions as well as the observed $m(\tauh,\tauh,\cPj,\cPj,\MET)$ and \ST spectra.
The last bin in the mass plot represents the yield for $m(\tauh,\tauh,\cPj,\cPj,\MET) > 2.25$\TeV,
while the last bin in the \ST plot represents the yield for $\ST > 1$\TeV (i.e. these bins include the overflow).
The observed yield is 14 events, while the predicted background yield is $19.8 \pm 4.2$ events, with QCD multijet, $\ttbar$,
and $\Z\to\tau\tau$ composing 76.3\%, 12.6\%, and 6.6\% of the total respectively (see Table~\ref{table:expectations}).
The simulated distributions corresponding to signal hypotheses with $m(\PW_\mathrm{R})=2.2$\TeV and $m(\mathrm{LQ})=0.8$\TeV
are also shown for comparison.
The observed $m(\tauh,\tauh,\cPj,\cPj,\MET)$ and \ST distributions do not reveal evidence for either
$\PW_\mathrm{R}\to\tau \cN_{\tau}\to\tau\tau \cPj\cPj$ or for LQ $\to\tau\PQb$ production.

The exclusion limit is calculated by using the distributions of $m(\tauh,\tauh,\cPj, \cPj, \MET)$
in the LRSM interpretation, or of \ST for the LQ interpretation, to construct the Poisson likelihood and to compute the 95\% Confidence Level (CL) upper
limit on the signal cross section $\sigma$ using the modified frequentist CL$_\mathrm{s}$ method \cite{Junk:1999kv, Read:2002hq}.
Systematic uncertainties are represented by nuisance parameters, assuming a gamma or log-normal prior for
normalization parameters, and Gaussian priors for shape uncertainties.

\begin{figure}[tbh!]
  \centering
    \includegraphics[width=0.48\textwidth]{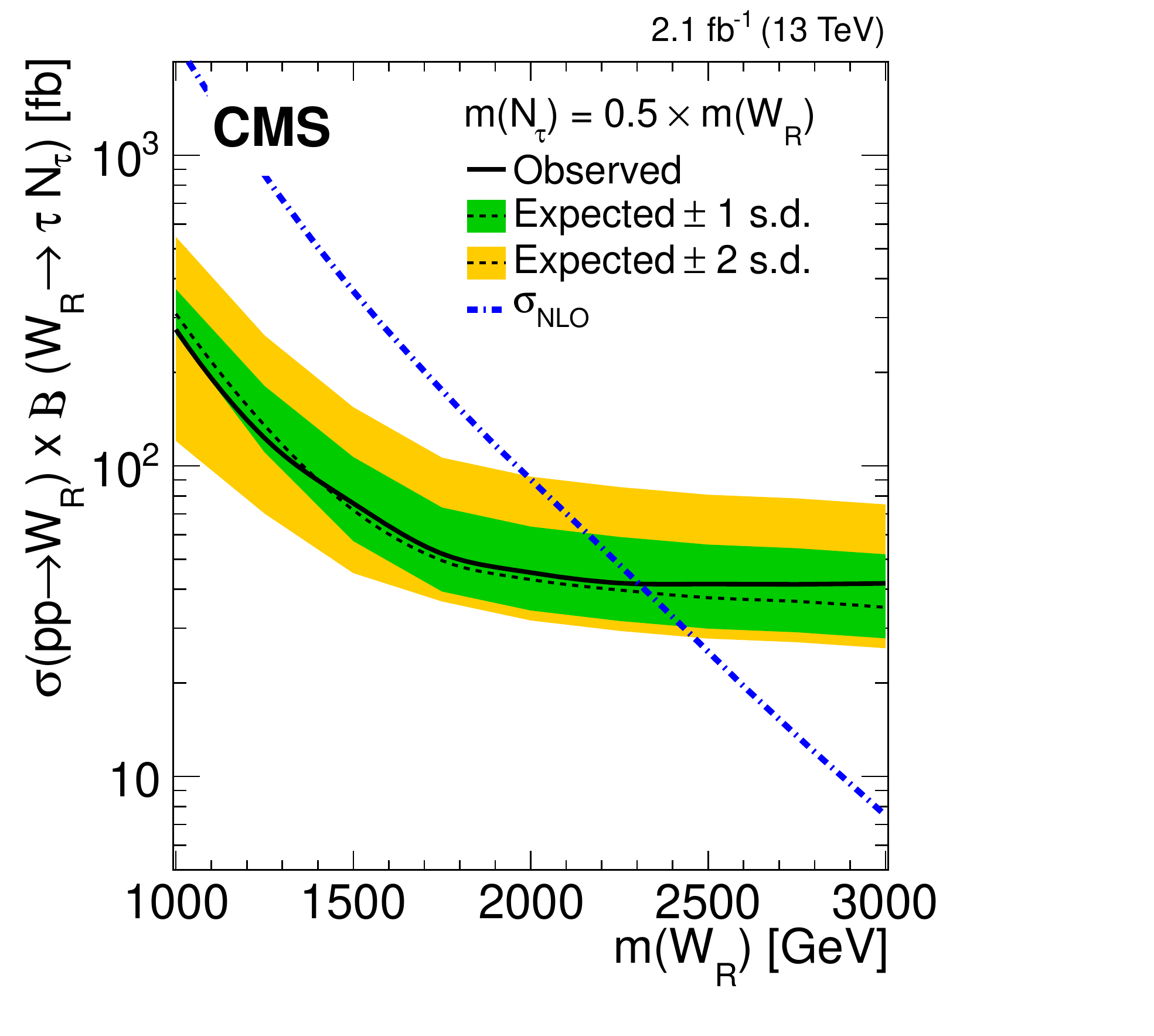}
    \includegraphics[width=0.48\textwidth]{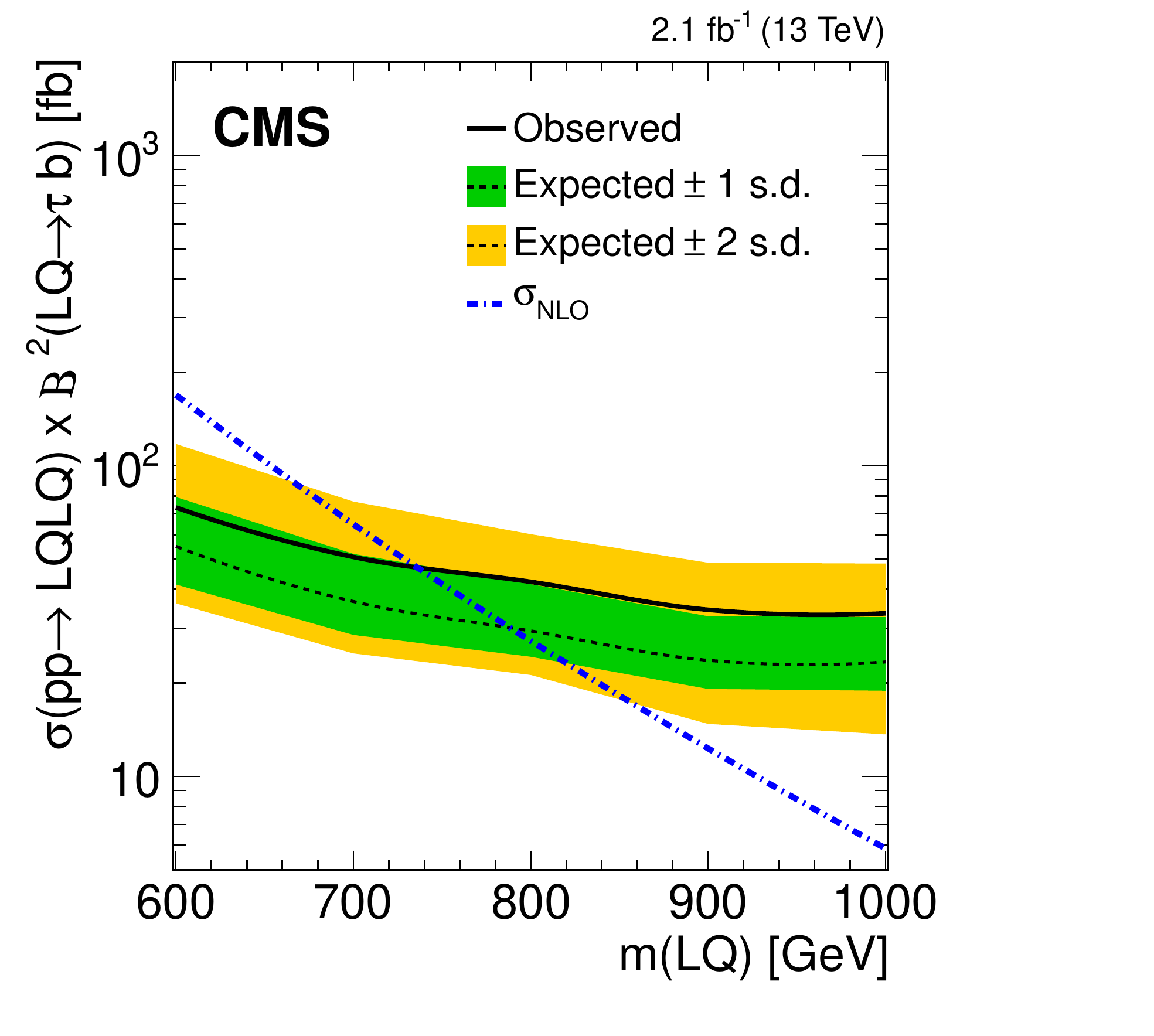}
  \caption{ Left: Expected and observed limits at 95\% CL on the product of $\PW_\mathrm{R}$ boson production cross section and
branching fraction of the $\PW_\mathrm{R} \to \tau \cN_{\tau}$ decay for $m(\cN_{\tau}) = m(\PW_\mathrm{R})/2$,
as functions of $m(\PW_\mathrm{R})$ mass.
Right: Expected and observed limits, at 95\% CL, on the product of the LQ pair production cross section and the branching fraction squared of the
$\mathrm{LQ} \to \tau\PQb$ decay, as functions of LQ mass. The bands around the expected limits represent the one and two standard deviation uncertainties
obtained using a large sample of pseudo-experiments based on the background-only hypothesis, for each bin of the mass and
\ST distributions. The dot-dashed blue line corresponds to the theoretical signal cross section at NLO, which assumes only
$\cN_{\tau}$ flavor contributes to the $\PW_\mathrm{R}$ boson decay width.}
\label{fig:OneDLimits}
\end{figure}

Figure~\ref{fig:OneDLimits} shows the expected and observed limits as well as the theoretical
cross sections as functions of $m(\PW_\mathrm{R})$ and $m(\mathrm{LQ})$.
For heavy neutrino models with strict left-right symmetry, and with the assumption that only
the $\cN_{\tau}$ flavor contributes to the $\PW_\mathrm{R}$ boson decay width, $\PW_\mathrm{R}$ boson masses below 2.3\TeV
are excluded at 95\% CL, assuming the $\cN_{\tau}$ mass is $0.5 \,m(\PW_\mathrm{R})$. The heavy-neutrino limits depend on the $\cN_{\tau}$ mass. 
For example, a scenario with $x = m(\cN_{\tau}) / m(\PW_\mathrm{R})$ $= 0.1$ (0.25) yields significantly lower average
jet and subleading $\tauh$ \pt than the $x = 0.5$ mass assumption, and the acceptance is lower by a factor of about 16 (3) for
$m(\PW_\mathrm{R}) = 1.0$\TeV and about 6.0 (1.9) for $m(\PW_\mathrm{R}) = 2.7$\TeV. On the other hand, the $x = 0.75$ scenario produces similar or larger average \pt for the jet and the $\tauh$ 
than the $x = 0.5$ mass assumption, yielding an event acceptance that is up to 10\% larger. 

\begin{figure}[tbh!]
  \centering
    \includegraphics[width=0.6\textwidth]{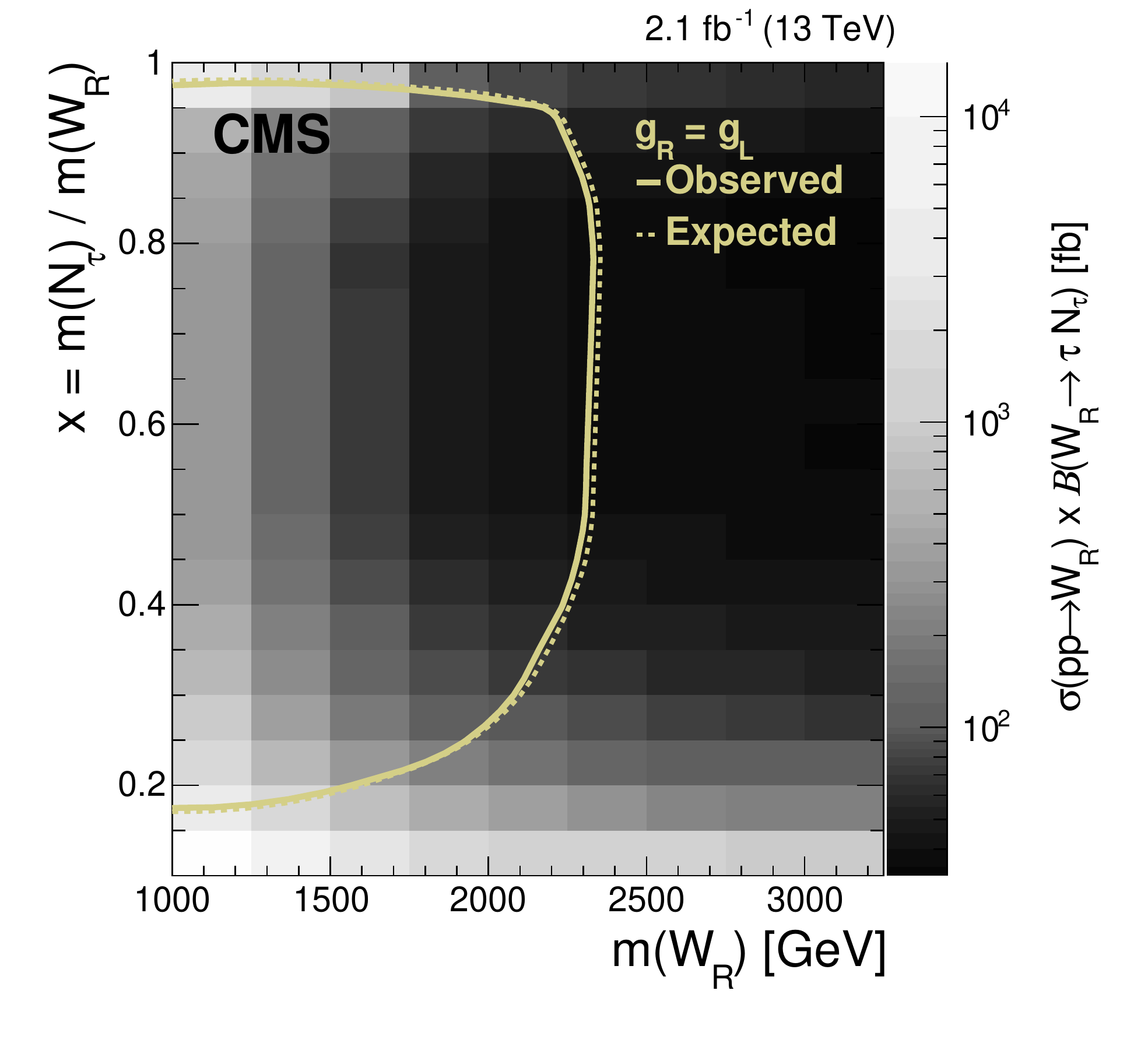}
  \caption{Exclusion bounds at 95\% CL as a function of $x = m(\cN_{\tau}) / m(\PW_\mathrm{R})$ and $m(\PW_\mathrm{R})$.
The color axis corresponds to the observed upper limit on the product of the cross section $\sigma(\Pp\Pp \to \PW_\mathrm{R})$
and the branching fraction $\mathcal{B}(\PW_\mathrm{R} \to \tau \cN_{\tau})$.  The curves indicate observed and
expected exclusion (left of the curve) for a model that assumes that the gauge couplings
associated with the right and left SU(2) groups are equal.}
\label{fig:2DSignalRegionPlot}
\end{figure}

Figure~ \ref{fig:2DSignalRegionPlot} shows the 95\% CL upper limits on the product of the production cross section and the branching fraction,
as a function of $m(\PW_\mathrm{R})$ and $x$. The signal acceptance and mass shape are evaluated for
each \{$m(\PW_\mathrm{R}), x$\} combination in Fig.~\ref{fig:2DSignalRegionPlot} and used in the limit calculation procedure described above.
Masses below $m(\PW_\mathrm{R})=2.35$ (1.63)\TeV are excluded at 95\% CL, assuming the $\cN_{\tau}$ mass is 0.8 (0.2) times the mass of $\PW_\mathrm{R}$ boson.

For the leptoquark interpretation using \ST as the final fit variable, LQ masses below 740\GeV are excluded at 95\% CL,
compared with expected limit of 790\GeV. The results of this search can also be applied to other models that predict a similar dilepton plus dijet final state, for 
example to the model with sterile right-handed neutrinos described in Ref.~\cite{SterileNeutrinos}.

\begin{table*}[htbp]
  \topcaption{Numbers of observed events in data and estimated background and signal rates in the signal region. The expected numbers of events for the
$\PW_\mathrm{R}$ signal samples assume $m(\cN_{\tau}) = m(\PW_\mathrm{R})/2$.}
  \centering{
  \begin{tabular}{ l | c }\hline
    Process  & Prediction \\ [0.5ex] \hline
    DY+jets & $1.3 \pm 0.5$    \\
    \PW{}+jets & $0.9 \pm 0.4$    \\
    $\ttbar$ & $2.5 \pm 0.9$    \\
    Multijet & $15.1 \pm 4.1$  \\
  \hline
    Total           & $19.8 \pm 4.2$    \\
  \hline
    Observed        & $14$                \\
  \hline
    $m(\PW_\mathrm{R})=1.0$\TeV & $61.1 \pm 1.5$    \\
    $m(\PW_\mathrm{R})=2.7$\TeV & $1.60 \pm 0.02$    \\
    $m(\mathrm{LQ})=0.6$\TeV & $14.7 \pm 0.3$    \\
    $m(\mathrm{LQ})=1.0$\TeV & $0.80 \pm 0.01$    \\
  \hline
  \end{tabular}
  }
  \label{table:expectations}
\end{table*}

\section{Summary}

A search is performed for physics beyond the standard model in events with two energetic $\tau$ leptons, two energetic jets, and large transverse
momentum imbalance, using a data sample corresponding to an integrated luminosity of 2.1\fbinv collected with the CMS detector in proton-proton collisions at $\sqrt{s}=13$\TeV.
The search focuses on two models: (1) production of heavy right-handed  third-generation neutrinos, $\cN_{\tau}$,
and right-handed $\PW_\mathrm{R}$ bosons that arise in the left-right symmetric extensions of the standard model, where the $\PW_\mathrm{R}$ decay chain results
in a pair of high-\pt $\tau$ leptons; (2)
pair production of third-generation scalar leptoquarks in the $\tau\tau\PQb\PQb$ channel.
The observed $m(\tauh,\tauh,\cPj,\cPj,\MET)$ and \ST distributions do not reveal any evidence of signals compatible with these scenarios.
Assuming that only the $\cN_{\tau}$ flavor contributes significantly to the $\PW_\mathrm{R}$ decay width, $\PW_\mathrm{R}$
masses below 2.35\,(1.63)\TeV are excluded at 95\% confidence level, assuming the $\cN_{\tau}$ mass is 0.8\,(0.2) times the mass of $\PW_\mathrm{R}$ boson. This analysis represents the first search for $N_{\tau}$ at the LHC and is also the first to focus on pair production of third-generation scalar
leptoquarks using the $\tauh\tauh\PQb\PQb$ final state.
Leptoquarks with a mass less than 740\GeV are excluded at 95\% confidence level, to be compared with an expected mass limit of 790\GeV.
This result equals the most stringent previous limit obtained in the $\tau_{l}\tauh\PQb\PQb$ final state, set by CMS using 19.5\fbinv of data recorded
at 8\TeV~\cite{LQ3CMS}.
This is the first search for third-generation leptoquarks in the $\tauh\tauh\PQb\PQb$ channel.

\begin{acknowledgments}
We congratulate our colleagues in the CERN accelerator departments for the excellent performance of the LHC and thank the technical and administrative staffs at CERN and at other CMS institutes for their contributions to the success of the CMS effort. In addition, we gratefully acknowledge the computing centers and personnel of the Worldwide LHC Computing Grid for delivering so effectively the computing infrastructure essential to our analyses. Finally, we acknowledge the enduring support for the construction and operation of the LHC and the CMS detector provided by the following funding agencies: BMWFW and FWF (Austria); FNRS and FWO (Belgium); CNPq, CAPES, FAPERJ, and FAPESP (Brazil); MES (Bulgaria); CERN; CAS, MoST, and NSFC (China); COLCIENCIAS (Colombia); MSES and CSF (Croatia); RPF (Cyprus); SENESCYT (Ecuador); MoER, ERC IUT, and ERDF (Estonia); Academy of Finland, MEC, and HIP (Finland); CEA and CNRS/IN2P3 (France); BMBF, DFG, and HGF (Germany); GSRT (Greece); OTKA and NIH (Hungary); DAE and DST (India); IPM (Iran); SFI (Ireland); INFN (Italy); MSIP and NRF (Republic of Korea); LAS (Lithuania); MOE and UM (Malaysia); BUAP, CINVESTAV, CONACYT, LNS, SEP, and UASLP-FAI (Mexico); MBIE (New Zealand); PAEC (Pakistan); MSHE and NSC (Poland); FCT (Portugal); JINR (Dubna); MON, RosAtom, RAS, and RFBR (Russia); MESTD (Serbia); SEIDI and CPAN (Spain); Swiss Funding Agencies (Switzerland); MST (Taipei); ThEPCenter, IPST, STAR, and NSTDA (Thailand); TUBITAK and TAEK (Turkey); NASU and SFFR (Ukraine); STFC (United Kingdom); DOE and NSF (USA).

\hyphenation{Rachada-pisek} Individuals have received support from the Marie-Curie program and the European Research Council and EPLANET (European Union); the Leventis Foundation; the A. P. Sloan Foundation; the Alexander von Humboldt Foundation; the Belgian Federal Science Policy Office; the Fonds pour la Formation \`a la Recherche dans l'Industrie et dans l'Agriculture (FRIA-Belgium); the Agentschap voor Innovatie door Wetenschap en Technologie (IWT-Belgium); the Ministry of Education, Youth and Sports (MEYS) of the Czech Republic; the Council of Science and Industrial Research, India; the HOMING PLUS program of the Foundation for Polish Science, cofinanced from European Union, Regional Development Fund, the Mobility Plus program of the Ministry of Science and Higher Education, the National Science Center (Poland), contracts Harmonia 2014/14/M/ST2/00428, Opus 2014/13/B/ST2/02543, 2014/15/B/ST2/03998, and 2015/19/B/ST2/02861, Sonata-bis 2012/07/E/ST2/01406; the Thalis and Aristeia programs cofinanced by EU-ESF and the Greek NSRF; the National Priorities Research Program by Qatar National Research Fund; the Programa Clar\'in-COFUND del Principado de Asturias; the Rachadapisek Sompot Fund for Postdoctoral Fellowship, Chulalongkorn University and the Chulalongkorn Academic into Its 2nd Century Project Advancement Project (Thailand); and the Welch Foundation, contract C-1845.
\end{acknowledgments}
\clearpage
\bibliography{auto_generated}

\cleardoublepage \appendix\section{The CMS Collaboration \label{app:collab}}\begin{sloppypar}\hyphenpenalty=5000\widowpenalty=500\clubpenalty=5000\textbf{Yerevan Physics Institute,  Yerevan,  Armenia}\\*[0pt]
V.~Khachatryan, A.M.~Sirunyan, A.~Tumasyan
\vskip\cmsinstskip
\textbf{Institut f\"{u}r Hochenergiephysik,  Wien,  Austria}\\*[0pt]
W.~Adam, E.~Asilar, T.~Bergauer, J.~Brandstetter, E.~Brondolin, M.~Dragicevic, J.~Er\"{o}, M.~Flechl, M.~Friedl, R.~Fr\"{u}hwirth\cmsAuthorMark{1}, V.M.~Ghete, C.~Hartl, N.~H\"{o}rmann, J.~Hrubec, M.~Jeitler\cmsAuthorMark{1}, A.~K\"{o}nig, I.~Kr\"{a}tschmer, D.~Liko, T.~Matsushita, I.~Mikulec, D.~Rabady, N.~Rad, B.~Rahbaran, H.~Rohringer, J.~Schieck\cmsAuthorMark{1}, J.~Strauss, W.~Waltenberger, C.-E.~Wulz\cmsAuthorMark{1}
% ...
\vskip\cmsinstskip
% ...
1:~~Also at Vienna University of Technology, Vienna, Austria\\
% ...

\end{sloppypar}
\end{document}